\documentclass[%
 aip,
 jap,%
 amsmath,amssymb,
preprint,%
]{revtex4-1}

\usepackage{graphicx}
\usepackage{dcolumn}
\usepackage{bm}
\usepackage{epstopdf}
\usepackage{xcolor}
\usepackage{multirow}

\begin{document}


\title[Sample title]{Effect of edge defects on band structure of zigzag graphene nanoribbons}

\author{Payal Wadhwa}
 \affiliation{Department of Physics, Indian Institute of Technology Ropar, Nangal Road, Rupnagar, Punjab - 140001,\ India}
\author{Shailesh Kumar}%
\affiliation{Manufacturing Flagship, CSIRO, Lindfield West, New South Wales 2070, Australia} 
\affiliation{School of Chemistry, Physics and Mechanical Engineering, Queensland University of Technology, Brisbane, Queensland 4000,\ Australia}
\author{T.J. Dhilip Kumar}
\affiliation{Department of Chemistry,\ Indian Institute of Technology Ropar,\ Rupnagar,\ Punjab -\ 140001,\ India}

\author{Alok Shukla}
\affiliation{Department of Physics, Indian Institute of Technology Bombay, Powai, Mumbai - 400076,\ India}
\affiliation{Department of Physics, Bennett University, Plot No 8-11, TechZone II, Greater Noida, Uttar Pradesh - 201310,\ India}

\author{Rakesh Kumar}
\email{rakesh@iitrpr.ac.in}
\affiliation{Department of Physics, Indian Institute of Technology Ropar, Nangal Road, Rupnagar, Punjab - 140001, India}

\date{\today}

\begin{abstract}
In this article,\ we report band structure studies of zigzag graphene nanoribbons (ZGNRs) on introducing defects ({\it $sp^{3}$} hybridized carbon atoms) in different concentrations at edges by varying the ratio of {\it $sp^{3}$} to {\it $sp^{2}$} hybridized carbon atoms.\ On the basis of theoretical analyses,\ band gap values of ZGNRs are found to be strongly dependent on relative arrangement of {\it $sp^{3}$} to {\it $sp^{2}$} hybridized carbon atoms at the edges for a defect concentration;\ so the findings would greatly help in understanding band gap of nanoribbons for their electronic applications.

\end{abstract}

\maketitle

\section{\label{sec:level1}Introduction}

Graphene,\ the first two dimensional material,\ was isolated for the first time in 2004 by A.K.\ Geim group at the University of Manchester\cite{PNAS,Geim}.\ Very soon,\ it attracted great attention from the scientific community for its extraordinary physical properties\cite{Nano,Science, Solid,Lee,PRB,Nanotech},\ but its zero bandgap turned out to be its biggest weakness, and remains the main bottlencek against its applications in electronics.\ Though,\ band structure calculations of graphene nanoribbons\ (GNRs) carried out even before discovery of graphene show metallic nature for ZGNRs, while semiconducting for armchair GNRs\cite{Nakada}.\ It was analyzed that formation of edge states at Fermi level from the edged carbon atoms in ZGNRs is responsible for its metallic nature,\ but possibility of a band gap in AGNRs prompted experimentalists to explore for a band gap by fabricating GNRs.\ However,\ the experiments on GNRs fabricated into mutually perpendicular crystallographic orientations by etching a single graphene sheet using electron beam lithography followed by oxygen plasma etching process showed band gaps in both zigzag and armchair crystallographic orientations \cite{Han},\ in contrast to the theoretical predictions\cite{Nakada,Waka,Son}.\ Thereafter, several theoretical attempts have been carried out so far to explain the experimental observations of ZGNRs in terms of edge defects,\ but have been restricted primarily to {\it $sp^{2}$} or {\it $sp^{3}$} hybridized carbon atoms at the edges\cite{Kusa,Yu,Cho}.\ It is now known that {\it $sp^{2}$} hybridization of carbon atoms at the edges on passivation with oxygen or hydrogen in ZGNRs leads to metallic behavior\cite{Cho,Nature,Pisani,APL},\ while {\it $sp^{3}$} hybridization leads to a small band gap value\cite{Cho,Rama}.\ It is to be noted that GNRs fabricated using oxygen plasma etching process may have some defective sites formed at the edges;\ motivated us to investigate band structures of ZGNRs on introducing defects at the edges.\ For studies,\ defect concentration is varied by changing the ratio of {\it $sp^{3}$} to {\it $sp^{2}$} hybridized carbon atoms at the edges as\ (a)\ 1:9,\ (b)\ 1:5,\ and (c)\ 1:3 for oxygen passivated edges of ZGNRs. \\

Band structure calculations have been performed using density functional theory as implemented in Vienna \textit{ab initio} simulation package (VASP)\cite{vasp}.\ Exchange correlational functional of Generalized gradient approximation (GGA)\cite{GGA}\ is used with electron-ion interactions as in projected augmented wave (PAW)\cite{PAW} formalism.\ A vacuum layer of at least 15 {\AA} is used to avoid interlayer interactions.\ The cut-off energy of 450 eV is used for theoretical band structure calculations.\ Before performing band structure calculations, system is relaxed until force on each atom is less than 0.001 eV.${\AA}^{-1}$.\ Monkhorst-Pack formalism is used for k-space sampling and k-mesh of size  $25\times1\times1$ is used for band structure calculations.\ Theoretical band structure calculations for defect studies of oxygen passivated ZGNRs are reported for non-magnetic ground states,\ because even on performing spin-polarized calculations, their ground states were found to be nonmagnetic (supplementary material I).\

\section{\label{sec:level1}Results and Discussion}

\begin{figure}[h]
\centering
  \includegraphics[height=3cm]{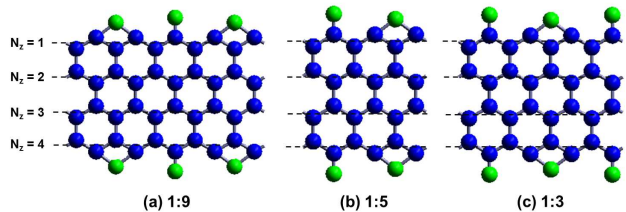}
  \caption{ZGNRs supercells with different defects concentration at edges for $N_{z} = 4$ on varying the ratio of {\it $sp^{3}$} to {\it $sp^{2}$} hybridized carbon atoms as (a) 1:9, (b) 1:5, and (c) 1:3. Blue and green spheres represent carbon and oxygen atoms, respectively.}
  \label{fgr:str1}
\end{figure}

For investigation of edge defects on band structure of ZGNRs,\ we consider three concentrations of defects.\ Edge defects are created by replacing some of the {\it $sp^{2}$} hybridized carbon atoms at edges by {\it $sp^{3}$} hybridized carbon atoms.\ Normally,\ GNRs are fabricated using oxygen plasma etching process,\ therefore oxygen have been used for passivation at the edges of ZGNRs.\ Defect concentration is varied by changing ratio of {\it $sp^{3}$} to {\it $sp^{2}$} hybridized carbon atoms at the edges for three cases as (a)\ 1:9,\ (b)\ 1:5,\ and (c)\ 1:3.\ For comparative studies,\ defects at edges are arranged in a symmetrical way w.r.t.\ each other as shown for a typical width of $N_{z} = 4$ in Figure \ref{fgr:str1}. 

\begin{figure}[h]
\centering
  \includegraphics[height=4.5cm]{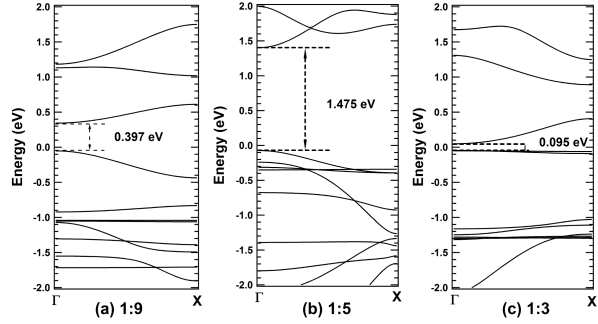}
  \caption{Band structure plots of 4-ZGNRs by varying the ratio of {\it $sp^{3}$} to {\it $sp^{2}$} hybridized carbon atoms at the edges as (a) 1:9, (b) 1:5, and (c) 1:3.}
  \label{fgr:str2}
\end{figure}

Band structure calculations are carried out for all the three cases of edge defects varying width from $N_{z} = 3$ to $N_{z} = 18$.\ Since,\ nanoribbons are quasi one-dimensional,\ therefore their band structures are plotted from $\Gamma$ to X point.\ Typical band structure plot of 4-ZGNRs for the three cases of defect concentrations are shown in Figure \ref{fgr:str2}. 

For different defect concentrations at edges,\ band gap values as a function of width are listed in Table \ref{tab:1}.\ For case (a) 1:9,\ semiconducting behavior is observed for ZGNRs at lower width $N_{z} \leq 6$, but becomes metallic at higher width $N_{z} > 6$.\ On increasing the edge defect concentration [case (b) 1:5] for a given width,\ band gap value increases as expected.\ In case (b),\ GNRs remains semiconducting for higher width in contrast to lower defect concentration of case (a).\ However,\ on further increasing the defect concentration [case (c) 1:3],\ surprisingly semiconducting behavior is observed only for $N_{z} = 3$ and 4,\ and becomes metallic for higher width of ZGNRs similar to that of case (a).\ In sharp contrast,\ the band gap values are also found to be lower than those of case (b),\ and even case (a). \\

\begin{table}[]
\centering
\caption{Band gap values as a function of width for different concentration of edge defects}
\label{tab:1}
\begin{tabular}{|c|c|c|c|c|}
\hline
\multicolumn{1}{|l|}{\multirow{2}{*}{\textbf{S.No.}}} & \multirow{2}{*}{\textbf{\begin{tabular}[c]{@{}c@{}}Width\\ $N_z$ \\\end{tabular}}} & \multicolumn{3}{c|}{\textbf{Band gap (eV)}}                           \\ \cline{3-5} 
\multicolumn{1}{|l|}{}                                &                                                                                             & \textbf{Case (a) 1:9} & \textbf{Case (b) 1:5} & \textbf{Case (c) 1:3} \\ \hline
1.                                                    & 3                                                                                           & 0.296                 & 0.877                 & 0.187                 \\ \hline
2.                                                    & 4                                                                                           & 0.397                 & 1.475                 & 0.095                 \\ \hline
3.                                                    & 5                                                                                           & 0.068                 & 0.56                  & Semi-metallic         \\ \hline
4.                                                    & 6                                                                                           & 0.103                 & 1.035                 & Semi-metallic         \\ \hline
5.                                                    & 7                                                                                           & Semi-metallic         & 0.393                 & Semi-metallic         \\ \hline
6.                                                    & 8                                                                                           & Semi-metallic         & 0.792                 & Semi-metallic         \\ \hline
7.                                                    & 9                                                                                           & Semi-metallic         & 0.310                 & Semi-metallic         \\ \hline
8.                                                    & 10                                                                                          & Semi-metallic         & 0.641                 & Semi-metallic         \\ \hline
9.                                                    & 11                                                                                          & Semi-metallic         & 0.243                 & Semi-metallic         \\ \hline
10.                                                   & 12                                                                                          & Semi-metallic         & 0.539                 & Semi-metallic         \\ \hline
11.                                                   & 13                                                                                          & Semi-metallic         & 0.205                 & Semi-metallic         \\ \hline
12.                                                   & 14                                                                                          & Semi-metallic         & 0.464                 & Semi-metallic         \\ \hline
13.                                                   & 15                                                                                          & Semi-metallic         & 0.177                 & Semi-metallic         \\ \hline
14.                                                   & 16                                                                                          & Semi-metallic         & 0.407                 & Semi-metallic         \\ \hline
15.                                                   & 17                                                                                          & Semi-metallic         & 0.157                 & Semi-metallic         \\ \hline
16.                                                   & 18                                                                                          & Semi-metallic         & 0.363                 & Semi-metallic         \\ \hline
\end{tabular}
\end{table}

From the above results,\ it is observed that band gap increases with defect concentration,\ but on further increasing the defect concentration,\ band gap value decreases.\ It can be attributed to the fact that with increase in the defect concentration at edges of ZGNRs,\ band gap would increase due to localization of charges at the edges states\cite{Rakesh}.\ On further increase in the defect concentration,\ the band gap decreases possibly as a consequence of a change in the  arrangement of {\it $sp^{3}$} and {\it $sp^{2}$} hybridized carbon atoms at the edges,\ which can be correlated to the edge reconstruction in nanoribbons \cite{P,Rodrigues,Kim,Liang,Ping}.\ Since in all the three cases,\ the {\it $sp^{3}$} and {\it $sp^{2}$} hybridized carbon atoms are arranged in a symmetrical way at the edges,\ but the only difference is the relative arrangement of {\it $sp^{3}$} to {\it $sp^{2}$} hybridized carbon atoms.\ It suggests that in addition to defect concentration,\ the order in which defects are arranged at the edges may play an important role in the band gap values of ZGNRs.\ It motivated us to investigate the effect of relative arrangement of {\it $sp^{3}$} to {\it $sp^{2}$} hybridized carbon atoms at the edges on the band gap values.\\

\begin{figure}[h]
\centering
  \includegraphics[height=3.5cm]{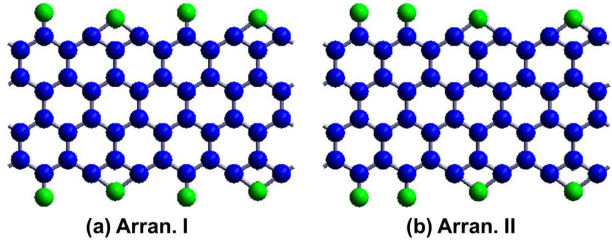}
  \caption{Two different symmetrical arrangements (Arran. I and Arran. II) of {\it $sp^{3}$} and {\it $sp^{2}$} hybridized carbon atoms at the edges for $N_{z} = 4$ ZGNRs supercells. Blue and green spheres represent carbon and oxygen atoms, respectively.}
  \label{fgr:str3}
\end{figure}

In order to investigate the effect of arrangements for a defect concentration,\ we consider different symmetrical arrangements of {\it $sp^{3}$} and {\it $sp^{2}$} hybridized carbon atoms at the edges,\ but with the same no.\ of atoms in the supercells for comparative analysis.\ Two different symmetrical arrangements (Arran.\ I and Arran.\ II) are considered for a typical width of $N_{z} = 4$ corresponding to a defect concentration of case (b) 1:5 as shown in Figure \ref{fgr:str3}. \\
  
\begin{figure}[h]
\centering
  \includegraphics[height=6cm]{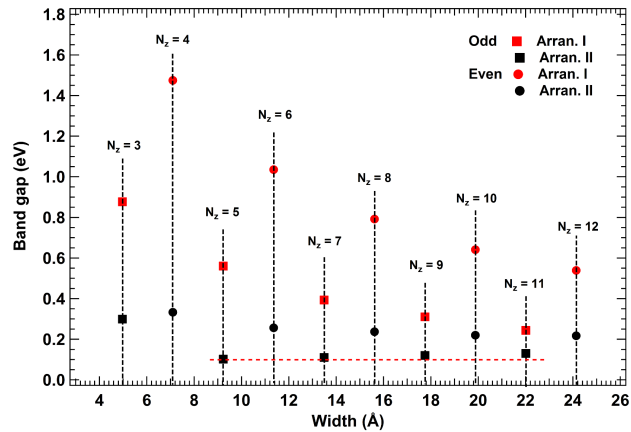}
  \caption{Band gap values for Arran.\ I and Arran.\ II are plotted as a function of width of $N_{z}$-ZGNRs. Red and black color corresponds to Arran.\ I and Arran.\ II, respectively. Horizontal dotted line (red) is drawn to show a small increase in band gap at higher width for Arran.\ II of odd $N_{z}$-ZGNRs.}
  \label{fgr:str4}
\end{figure}

Band structure calculations are carried out for $N_{z} = 3$ to 12.\ It is found that the band gap values are different for both the arrangements, Arran.\ I and Arran.\ II as shown in Figure \ref{fgr:str4}.\ It is to be noted that band gap values as a function of width are distributed into two classes corresponding to even $N_{z}$ and odd $N_{z}$,\ even though band gap decreases with width as expected  as a consequence of increase in the confinement length along the width.\ Consequently,\ difference in band gap values decreases with width between Arran.\ I and Arran.\ II for both even and odd $N_{z}$-ZGNRs.\ However,\ a small increase in band gap corresponding to Arran.\ II for odd $N_{z}$-ZGNRs is observed at higher width possibly as a consequence of intra-edge interactions.\ It can also be noted that band gap increases with width on moving from odd $N_{z}$ to even $N_{z}$ for the same arrangement,\ e.g,\ from $N_{z}= 5$ to $N_{z}= 6$.

\begin{figure}[h]
\centering
  \includegraphics[height=6cm]{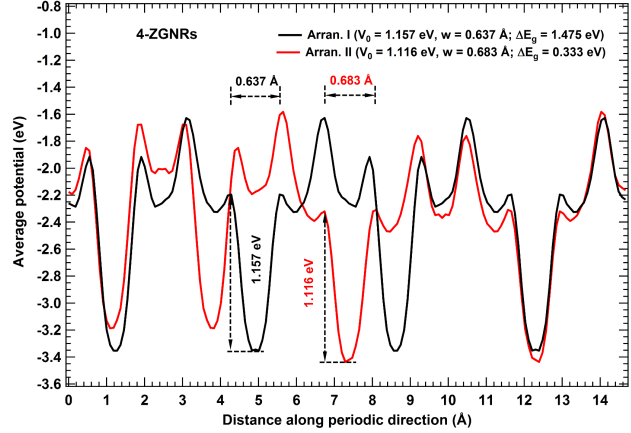}
  \caption{Average potential profile as a function of distance in periodic direction for 4-ZGNRs supercell. Potential profile for Arran.\ I and Arran.\ II are shown in black and red, respectively.}
  \label{fgr:str5}
\end{figure}

In order to investigate,\ how the arrangement affects band gap value,\ we consider a typical width of $N_{z} = 4$.\ Band gap calculated for Arran.\ I and Arran.\ II are 1.475 eV and 0.333 eV, respectively.\ For theoretical analyses,\ their one-dimensional potential profiles are plotted (Figure \ref{fgr:str5}) and the potential well at global minimum is considered for comparative analysis\cite{Deepika}.\ It can be seen from the Figure \ref{fgr:str5} that depth of the deepest potential well at global minimum ($V_{0}$) for Arran.\ I and Arran.\ II are 1.157 eV and 1.116 eV, respectively,\ while width of the potential well (w) at global minimum are 0.637 \AA \ and 0.683 \AA, respectively. 

From the analysis of multiple quantum well structures,\ different band gap values of  Arran.\ I and Arran.\ II are correlated with the depth and width of the potential well using the equation according to Kronig-Penney model\cite{KP}
\begin{equation}
\frac{A^{2}-B^{2}}{2AB}\sin(Aw)\sinh(Bw)+\cos(Aw)\cosh(Bw)=\cos(k.2w)\\
\end{equation}
where,
\begin{eqnarray}{\label{eqn:eqnad}}
B=\sqrt{\frac{2mV_{0}}{\hbar^{2}}-A^{2}}, 
\frac{\hbar^{2}A^{2}}{2m}=E+V_{0}>0
\end{eqnarray}
and 
\begin{eqnarray}{\label{eqn:eqnad}}
\frac{\hbar^{2}B^{2}}{2m}=-E<0
\end{eqnarray}

On solving equation (1) with the corresponding values of width and depth for Arran.\ I and Arran.\ II,\ the band gap estimated for Arran.\ I is found to be greater than that of Arran.\ II,\ which explains how a change in the arrangement affects the periodic potential profile to give different band gap value.\ For further analysis, ground state energy per atom calculated for supercells corresponding to Arran.\ I and Arran.\ II are - 8.453 eV and - 8.427 eV,\ respectively.\ Since their ground state energies differ by small amounts,\ therefore both the arrangements are almost equally probable.\ However on edge reconstructions, Arran.\ I may be favored over Arran.\ II,\ similar to the reconstruction patterns formed in transition metal dichalcogenides based nanoribbons\cite{Ping}.\ Similar analyses are also valid for odd $N_{z}$-ZGNRs (supplementary material II).\ Similarly on the basis of potential profile,\ higher band gap value of even $N_{z}$-ZGNRs than that of odd $N_{z}$-ZGNRs are explained but needs normalization of potential depth w.r.t.\ width for comparison\cite{Deepika} [supplementary material III].\ The studies indicate that a change in arrangement of defects ({\it $sp^{3}$} hybridized carbon atoms w.r.t.\ the {\it $sp^{2}$} hybridized carbon atoms) at the edges modifies the average potential profile in the periodic direction,\ which is responsible for a change in the band gap values.\\

\begin{figure}[h]
\centering
  \includegraphics[height=3.5cm]{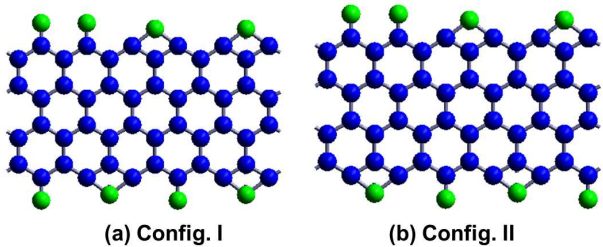}
  \caption{ZGNRs supercells for two different asymmetrical configurations at the edges for defect concentration corresponding to case (b) 1:5. Blue and green spheres represent carbon and oxygen atoms, respectively.}
  \label{fgr:str6}
\end{figure}

In order to investigate further the effect of arrangement on band gap values,\ the relative arrangement of {\it $sp^{3}$} to {\it $sp^{2}$} hybridized carbon atoms at the edges are considered also in asymmetrical way.\ Two different asymmetrical configurations (Config.\ I and Config.\ II) for a defect concentration of case (b) 1:5 for a typical width of $N_{z} = 4$ are shown in Figure \ref{fgr:str6}.  

Band gap calculated for  Config.\ I and Config.\ II are 0.207 eV and 0.473 eV, respectively;\ which is also verified on the basis of their one-dimensional potential profile as discussed earlier.\ The ground state energy per atom calculated for Config.\ I and Config.\ II are - 8.441 eV and - 8.447 eV, respectively.\ Since their ground state energies also differ by small amounts,\ therefore both the arrangements are almost equally probable.\ However on edge reconstructions,\ Config.\ II may be favored over Config.\ I.\ It further ensures that in addition to defect concentration,\ relative arrangements of {\it $sp^{3}$} to {\it $sp^{2}$} hybridized atoms at the edges play an important role in deciding the band gap of nanoribbons. 

\section{\label{sec:level1}Conclusion}
It is concluded that band gap of ZGNRs are  strongly affected by relative arrangement of defects ({\it $sp^{3}$} to {\it $sp^{2}$} hybridized carbon atoms) at the edges for a defect concentration,\ which is a consequence of a change in their one-dimensional potential profile.\ The investigations would help in understanding the band gap studies of nanoribbons for their electronic applications.\\

\section*{Supplementary Material}
See supplementary material for the analyses of different band gap values of Arran.\ I and Arran.\ II for odd $N_{z}$-ZGNRs and higher band gap of even $N_{z}$-ZGNRs than that of odd $N_{z}$-ZGNRs.

\section*{ACKNOWLEDGEMENTS}
The authors would like to acknowledge HPC facility at IIT Ropar and MHRD for funding.\\

\section*{\label{sec:level1}References}

\end{document}